\documentclass{article}

\usepackage{PRIMEarxiv}

\usepackage[utf8]{inputenc} 
\usepackage[T1]{fontenc}    
\usepackage{hyperref}       
\usepackage{url}            
\usepackage{booktabs}       
\usepackage{amsfonts}       
\usepackage{nicefrac}       
\usepackage{microtype}      
\usepackage{lipsum}
\usepackage{fancyhdr}       
\usepackage{graphicx}       
\graphicspath{{media/}}     

\title{Spiritus: An AI-Assisted Tool for Creating 2D Characters and Animations}

\author{
  Qirui Sun, Yunyi Ni, Teli Yuan, Jingjing Zhang, Fan Yang, Zhihao Yao, Haipeng Mi
  \thanks{
    Qirui Sun is with Tsinghua University and Kotoko AI, Beijing, China. 
    Yunyi Ni is with Nanyang Technological University, Singapore. 
    Teli Yuan is with Kotoko AI, Japan. 
    Jingjing Zhang is with Beijing Jiaotong University, Beijing, China. 
    Fan Yang is with University of Washington, Seattle, United States. 
    Zhihao Yao is with Tsinghua University, Beijing, China. 
    Haipeng Mi is with Tsinghua University, Beijing, China (corresponding author, e-mail: mhp@tsinghua.edu.cn).
  }
}

\begin{document}


\maketitle

\begin{abstract}
This research presents Spiritus, an AI-assisted creation tool designed to streamline 2D character animation creation while enhancing creative flexibility. By integrating natural language processing and diffusion models, users can efficiently transform natural language descriptions into personalized 2D characters and animations. The system employs automated segmentation, layered costume techniques, and dynamic mesh-skeleton binding solutions to support flexible adaptation of complex costumes and additional components. Spiritus further achieves real-time animation generation and efficient animation resource reuse between characters through the integration of BVH data and motion diffusion models. Experimental results demonstrate Spiritus's effectiveness in reducing technical barriers, enhancing creative freedom, and supporting resource universality. Future work will focus on optimizing user experience and further exploring the system's human-computer collaboration potential.
\end{abstract}


\section{INTRODUCTION}

\begin{figure}[ht]
  \includegraphics[width=\textwidth]{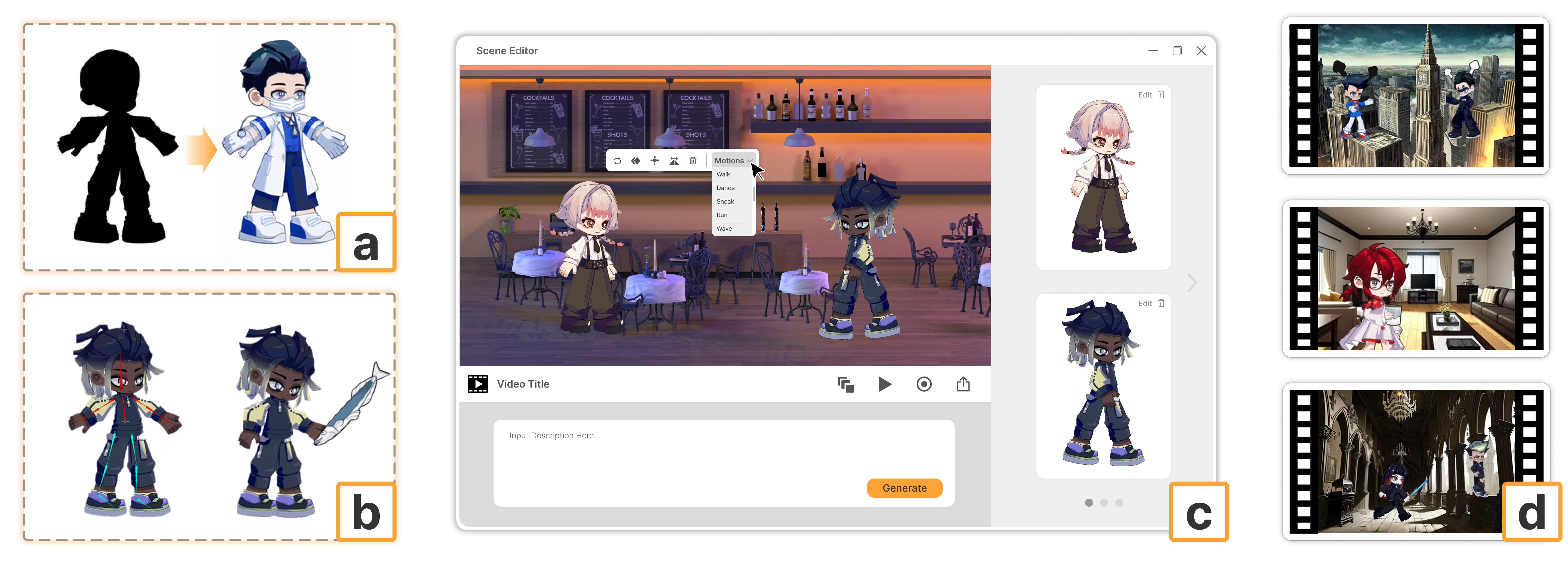}
  \caption{Spiritus is an AI-assisted creation tool that integrates natural language processing, image generation, and skeletal animation technologies. Spiritus aims to provide personalized creative support for creators focused on character and animation creation: (a) Character image generation, (b) Animation and Hand components, (c) Scene orchestration, (d) User works. }
  \label{fig:teaser}
\end{figure}

In recent years, with the rise of short-video platforms, games, and GIF memes, dynamic visual media has emerged as one of the predominant narrative forms in social media \cite{shutsko2020user,wang2019causes}. Animated skits featuring 2D character animations, as an emerging form of digital content, have rapidly gained prominence across global social media platforms \cite{li2023does,ojomo2021social}. These skits are characterized by brief, humorous, or satirical performances, demonstrating high shareability and creative flexibility. Unlike traditional animation forms, animated skits primarily utilize creator-developed IPs or game IPs as narrative subjects, offering creators greater freedom in visual styling and enabling the incorporation of exaggerated and surreal elements to enhance expressiveness \cite{farnham2011faceted}. As an innovative form of self-expression, animated skits particularly appeal to younger generations, enabling them to explore and express multiple identities through virtual characters \cite{harrell2012imagination}.

As AI-driven generative tools become increasingly prominent in content creation, particularly text-to-image \cite{mansimov2015generating} and text-to-video \cite{liu2024sora} technologies have equipped content creators with powerful visual content generation capabilities. These technologies have fostered innovations in dynamic narrative expression and visual creativity, exemplified by systems such as ID.8 by Antony et al., which demonstrates AI's potential in visual storytelling \cite{10.1145/3672277}, and the generative story ideation tools by Chung et al., which further expand the possibilities of narrative creation \cite{10.1145/3491102.3501819}. Additionally, Wolf et al. and Zhao et al. have achieved pioneering breakthroughs in character generation and parametric construction respectively \cite{wolf2017unsupervised,Zhao_2023_CVPR}.

Despite the new possibilities enabled by AI-integrated animation technologies, current creation tools and AI animation generation systems face significant limitations. On one hand, traditional tools (e.g., After Effects, Live2D, and Spine) require substantial professional expertise and time investment, limiting non-professional users' creative potential. On the other hand, AI-based animation generation methods, such as image generation approaches for motion sequence frames, while partially simplifying the production process, still face challenges in character consistency and generation process control. For instance, existing systems \cite{guo2023momask,guo2023sparsectrl,Guajardo} often struggle with maintaining consistent character styles in generated animation frames and provide limited post-generation editing capabilities. Furthermore, while some research utilizing skeleton animation retargeting methods \cite{Smith} has achieved automated sketch-to-animation conversion, fine-grained control over character dynamic generation remains inadequate.

To explore new workflows and methods addressing these limitations, we present Spiritus, an AI-assisted creation tool that integrates natural language processing, text-to-character generation, and skeleton animation technologies. Spiritus provides a comprehensive solution from text input to character generation and animation matching, supporting animated skit creation with low entry barriers and high creative freedom. Its key advantages include:
\begin{enumerate}
\item \textbf{Algorithm: }
Developed a hierarchical character generation model with semantic-based segmentation and unified rigging for stable character creation and animation.
\item \textbf{System: }
Created a web-based interactive platform supporting text-to-character generation and animation creation through text prompts and BVH mapping.
\item \textbf{Data: }
Enabled cross-platform character and animation workflow with import/export capabilities via Spine Runtime.
\end{enumerate}

\section{USER INTERFACE AND INTERACTION}

Spiritus is a web-based interactive system designed to help 2D character animation creators rapidly generate dynamic presentations of personalized creative stories with AI assistance. Spiritus consists of three core components: scene orchestration, character creation, and animation generation components, including creative support functionalities.

\subsection{Scene Orchestration Component}

\begin{figure}[ht]
    \centering
    \includegraphics[width=0.7\linewidth]{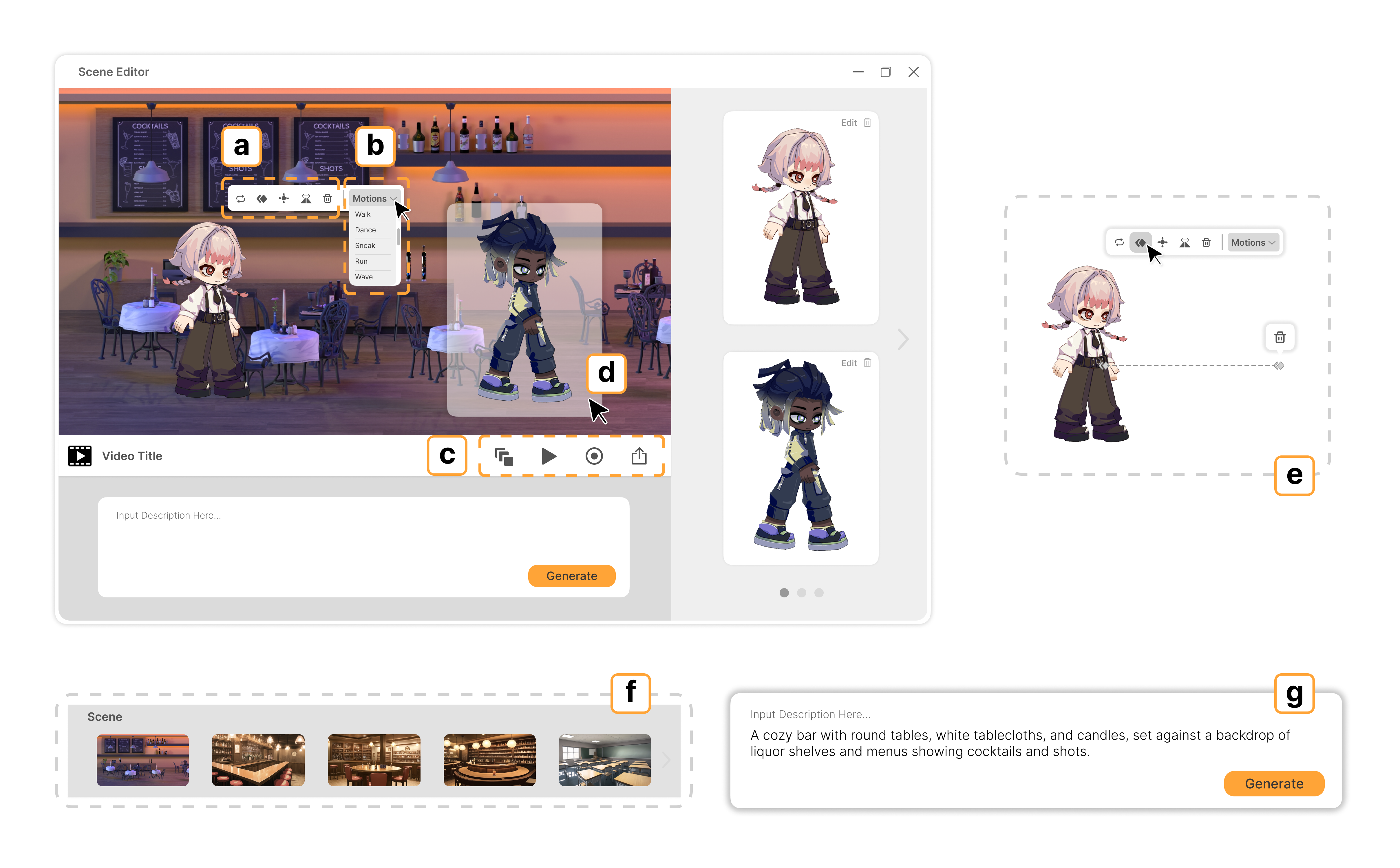}
    \caption{Scene orchestration component interface, editing tools include (a) loop/single play, keyframe, move, mirror, delete tools, (b) character animation selection, (c) scene switching, animation playback, recording, export tools; interaction methods include (d) dragging characters into the scene, (e) adding and deleting keyframe animation paths, (f) scene selection, (g) scene text input generation.}
    \label{fig:43}
\end{figure}

The scene creation and narrative module includes scene generation and creative storyboard tools. Users can input a short text describing their creative requirements, including story background, settings, and plot elements. They can generate scene images as stage backgrounds for their mini-theater creations through pure natural language descriptions, with the system generating environmental elements, objects, and atmosphere based on these descriptions (Figure \ref{fig:43}(g)). Users can select and save one or multiple satisfactory scenes for subsequent plot development (Figure \ref{fig:43}(f)). The scenes support dynamic character embedding, allowing creators to flexibly arrange scene narratives. Users can utilize scenes as storyboard tools for narrative creation, constructing plot and character interactions by designing and modifying elements within the virtual environment. This process not only enables creators to plan story development more intuitively but also stimulates new creativity and inspiration, enriching narrative techniques. After creating the main view of the complete scene animation, users can subsequently create one or more characters and place them in appropriate positions within the scene through drag-and-drop functionality.

\subsection{Character Creation Interface}

\begin{figure}[ht]
    \centering
    \includegraphics[width=1\linewidth]{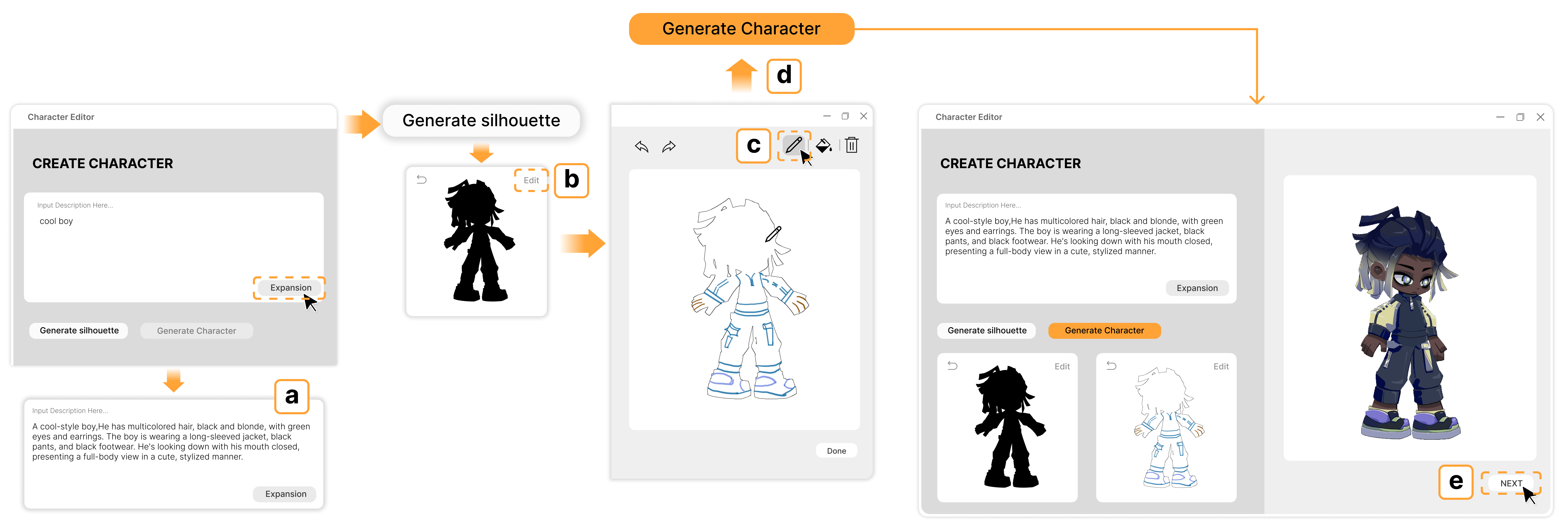}
    \caption{Character creation interface workflow, including: (a) system supplementary description, (b) editing texture details based on the outline, (c) texture drawing, (d) selecting character outline or editing texture outline for character generation, (e) generation complete. }
    \label{fig:41}
\end{figure}

In the scene interface, clicking on the default character opens the character generation interface, where creators can create characters through a combination of natural language descriptions and sketches. This character can represent the creator themselves or any narrative character.

Specifically, users first input natural language descriptions of the character's appearance (Figure \ref{fig:41}(a)). They can provide detailed descriptions of clothing style, facial expressions, outfit types, and even eyelash color, or simply specify gender. Users can also input a very simple character description and then opt for the system to generate more detailed character specifications, which they can later choose to adjust. For example:
User description: A cool boy
System Expansion: A chibi-style boy. He has multicolored hair, black and blonde, with green eyes and earrings. The boy is wearing a long-sleeved jacket, black pants, and black footwear. He's looking down with his mouth closed, presenting a full-body view in a cute, stylized manner.

The system also provides real-time preview and adjustment capabilities for generated characters, including feature modification and new alternative generations. Based on the user's text, the system generates a character silhouette (Figure \ref{fig:41}(b)). Users can choose to either directly generate a complete character image or add customized texture details through sketching (Figure \ref{fig:41}(d)). When users are not satisfied with the generated character, they can choose to return to either the silhouette generation stage or the sketching stage for modifications.

\subsection{Animation and Hand Component Interface}

\begin{figure}
    \centering
    \includegraphics[width=1\linewidth]{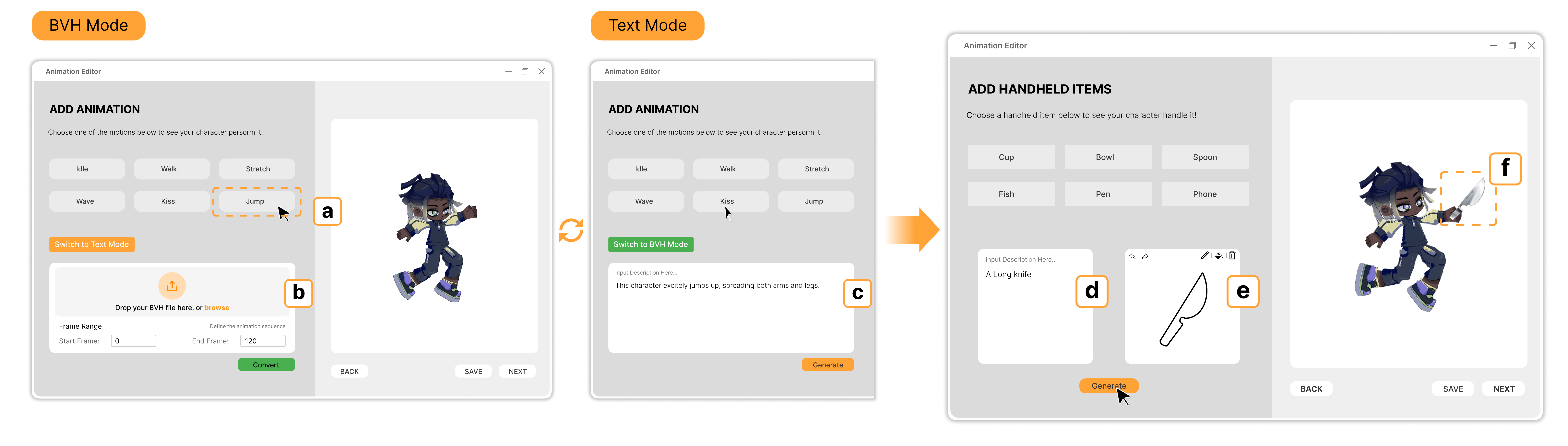}
    \caption{Animation editing and hand component interface. (a) Action selection, (b) BVH file conversion to character animation, (c) Text input for AI-generated animation, (d) Hand prop text input, (e) Hand prop sketching, (f) Example interface for hand extension component.}
    \label{fig:42}
\end{figure}

Based on the generated character, users can select basic animation actions from the animation library to preview and save effects (Figure \ref{fig:42}(a)). They can also input text descriptions of character actions or use BVH animation files to generate real-time animation effects that meet their requirements (Figure \ref{fig:42}(b,c)). Animation generation allows for very free expression of movements, such as jumping, bending, laughing heartily, showing anger, attacking, etc. We have pre-generated and created commonly used actions suggested in creator interviews, such as idle, waving, walking, running, and jumping, for creators to use quickly.

For professional users, they can download the generated animation files and dynamically adjust them in Spine software as needed until they achieve satisfactory action animations, ensuring creative expression and animation optimization throughout the creation process.

After selecting animations, each character has an extension component interface for their hands, allowing users to freely generate hand-connected components as extensions for character animation narratives (Figure \ref{fig:42}(d,e,f)).

\section{IMPLEMENTATION}

\begin{figure}
    \centering
    \includegraphics[width=0.8\linewidth]{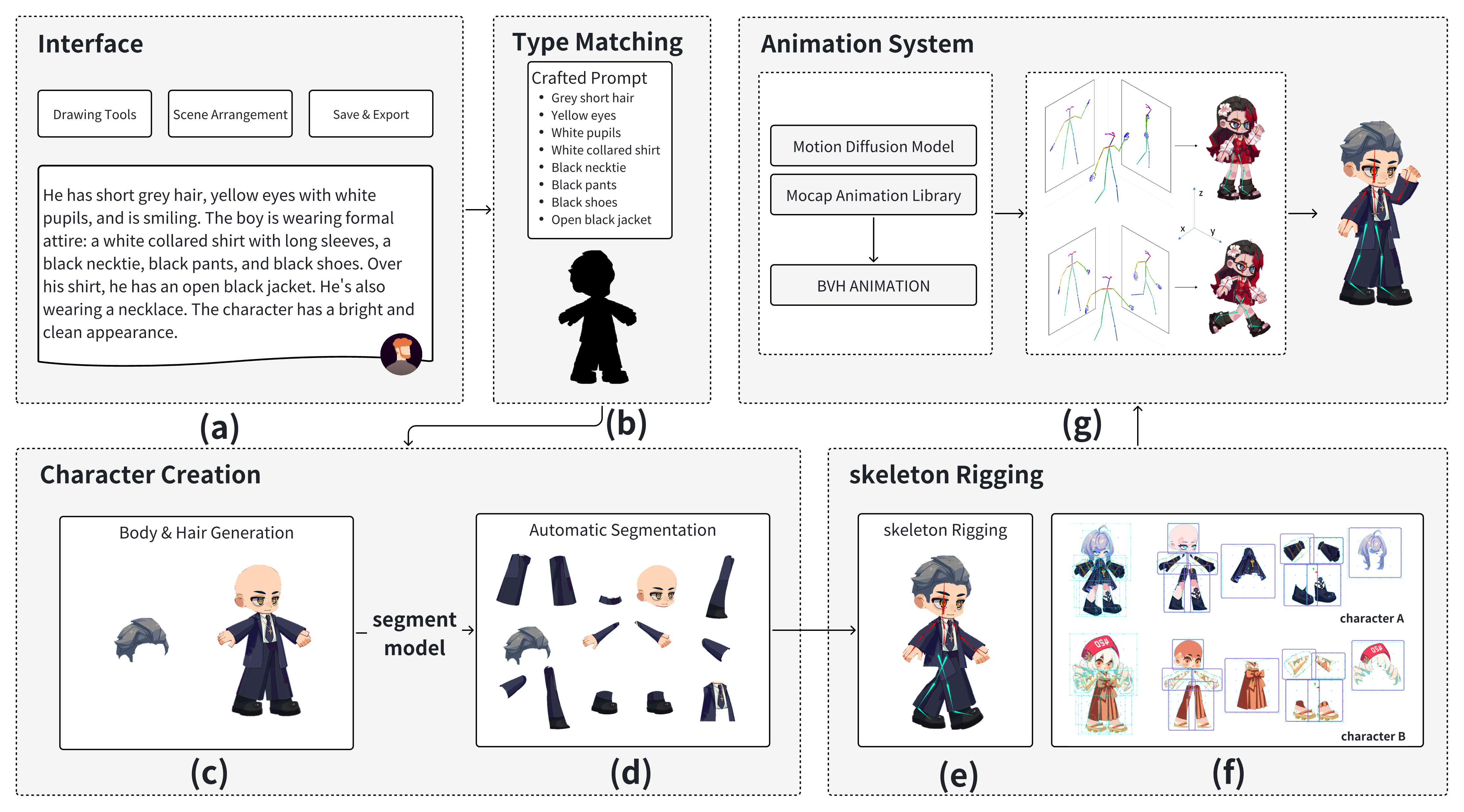}
    \caption{Implementation Framework }
    \label{fig:frame}
\end{figure}

Spiritus's implementation primarily consists of two parts: character generation and animation system generation. Through natural language processing, the system's core framework is shown in Figure \ref{fig:frame}, and Spiritus's automated workflow simplifies character creation and animation generation through several key steps.

\subsection{Image Generation and Segmentation}
First, the system uses the Stable Diffusion (SD) model to transform users' natural language descriptions into character images. It enhances input text through Large Language Models (LLM) and uses Named Entity Recognition (NER) to extract appearance-related information, matching it with a preset vocabulary library to generate silhouettes and serve as segmentation references. Users can choose from multiple design variants (Figure \ref{fig:frame}(a,b)).

Second, utilizing previously recorded character component semantics and the Segment Anything Model (SAM) \cite{kirillov2023segment}, the system segments character images into logical regions (such as head, limbs, clothing) as the foundation for layered animation (Figure \ref{fig:frame}(d)).

\subsection{Skeleton and Dynamic Mesh Binding}
Spiritus's animation binding system incorporates dynamic mesh and layering techniques to optimize the handling of complex costumes and accessories in 2D animation while enhancing the dynamic effects of AI-generated characters. To achieve freedom in costume generation and automatic binding, we reference Sun et al.'s research \cite{sun2024} and redesign a non-uniform mesh system that can deform with skeletal movement. Each mesh stores one type of additional component, ensuring different attachments automatically use corresponding meshes and can store various styles of accessories. This flexibility allows characters with identical attachments to display different visual effects (Figure \ref{fig:frame}(f)). A significant advantage of this approach is that all characters with different costumes can be driven by unified skeletal animation, laying the foundation for subsequent animation asset transfer between different characters.

\subsection{Animation Generation and Mapping}

Finally, inspired by Smith et al.'s work \cite{Smith}, the system implements skeleton animation generation based on BVH data mapping, adopting a hybrid perspective mapping method to adapt BVH to 2D character geometry (Figure \ref{fig:frame}(g)). By selecting optimal projection planes, 3D motion data is mapped to 2D space, with frontal projection for head and torso, and a combination of frontal and side projections for limbs, using Principal Component Analysis (PCA) to determine projection planes for each joint. The animation system can prioritize searching the BVH animation library for matching motions based on user input motion descriptions. Additionally, utilizing the Motion Diffusion Model (MDM) \cite{tevet2023human}, real-time 3D motion data can be generated and mapped to BVH format to support real-time animation options. However, due to the not fully stable nature of generated animations, this functionality is considered for future workflow exploration.

\section{CONCLUSION}

Spiritus is an AI-assisted tool for 2D character generation and animation creation that lowers technical barriers through automation. It transforms text descriptions into personalized 2D characters using natural language processing and image generation. The system enables dynamic expressions through automated segmentation, layering, and skeleton binding, while supporting costume adaptations. By leveraging BVH data and motion diffusion models, Spiritus enables instant animation generation and resource sharing.Future work will focus on flexible skeleton editing, diverse character generation, and enhanced creator interactions, aiming to inspire more creative animation production.


\bibliographystyle{unsrt}  
\bibliography{ref}  

\begin{thebibliography}{10}

\bibitem{shutsko2020user}
Aliaksandra Shutsko.
\newblock User-generated short video content in social media. a case study of tiktok.
\newblock In {\em Social Computing and Social Media. Participation, User Experience, Consumer Experience, and Applications of Social Computing: 12th International Conference, SCSM 2020, Held as Part of the 22nd HCI International Conference, HCII 2020, Copenhagen, Denmark, July 19--24, 2020, Proceedings, Part II 22}, pages 108--125. Springer, 2020.

\bibitem{wang2019causes}
Yu-Huan Wang, Tian-Jun Gu, and Shyang-Yuh Wang.
\newblock Causes and characteristics of short video platform internet community taking the tiktok short video application as an example.
\newblock In {\em 2019 IEEE International Conference on Consumer Electronics-Taiwan (ICCE-TW)}, pages 1--2. IEEE, 2019.

\bibitem{li2023does}
Yijin Li.
\newblock Why does gen z watch virtual streaming vtube anime videos with avatars on twitch?
\newblock {\em Online Media and Global Communication}, 2(3):379--403, 2023.

\bibitem{ojomo2021social}
Olusegun Ojomo and Oluwaseyi~Adewunmi Sodeinde.
\newblock Social media skits: Reshaping the entertainment experience of broadcast audience.
\newblock {\em Sage Open}, 11(3):21582440211032176, 2021.

\bibitem{farnham2011faceted}
Shelly~D Farnham and Elizabeth~F Churchill.
\newblock Faceted identity, faceted lives: social and technical issues with being yourself online.
\newblock In {\em Proceedings of the ACM 2011 conference on Computer supported cooperative work}, pages 359--368, 2011.

\bibitem{harrell2012imagination}
D~Fox Harrell and S~Veeragoudar Harrell.
\newblock Imagination, computation, and self-expression: Situated character and avatar mediated identity.
\newblock {\em Leonardo electronic almanac}, 17(2), 2012.

\bibitem{mansimov2015generating}
Elman Mansimov, Emilio Parisotto, Jimmy~Lei Ba, and Ruslan Salakhutdinov.
\newblock Generating images from captions with attention.
\newblock {\em arXiv preprint arXiv:1511.02793}, 2015.

\bibitem{liu2024sora}
Yixin Liu, Kai Zhang, Yuan Li, Zhiling Yan, Chujie Gao, Ruoxi Chen, Zhengqing Yuan, Yue Huang, Hanchi Sun, Jianfeng Gao, et~al.
\newblock Sora: A review on background, technology, limitations, and opportunities of large vision models.
\newblock {\em arXiv preprint arXiv:2402.17177}, 2024.

\bibitem{10.1145/3672277}
Victor~Nikhil Antony and Chien-Ming Huang.
\newblock Id.8: Co-creating visual stories with generative ai.
\newblock {\em ACM Trans. Interact. Intell. Syst.}, 14(3), aug 2024.

\bibitem{10.1145/3491102.3501819}
John Joon~Young Chung, Wooseok Kim, Kang~Min Yoo, Hwaran Lee, Eytan Adar, and Minsuk Chang.
\newblock Talebrush: Sketching stories with generative pretrained language models.
\newblock In {\em Proceedings of the 2022 CHI Conference on Human Factors in Computing Systems}, CHI '22, New York, NY, USA, 2022. Association for Computing Machinery.

\bibitem{wolf2017unsupervised}
Lior Wolf, Yaniv Taigman, and Adam Polyak.
\newblock Unsupervised creation of parameterized avatars.
\newblock In {\em 2017 IEEE International Conference on Computer Vision (ICCV)}, pages 1539--1547, 2017.

\bibitem{Zhao_2023_CVPR}
Rui Zhao, Wei Li, Zhipeng Hu, Lincheng Li, Zhengxia Zou, Zhenwei Shi, and Changjie Fan.
\newblock Zero-shot text-to-parameter translation for game character auto-creation.
\newblock In {\em 2023 IEEE/CVF Conference on Computer Vision and Pattern Recognition (CVPR)}, pages 21013--21023, 2023.

\bibitem{guo2023momask}
Chuan Guo, Yuxuan Mu, Muhammad~Gohar Javed, Sen Wang, and Li~Cheng.
\newblock Momask: Generative masked modeling of 3d human motions.
\newblock {\em arXiv preprint arXiv:2312.00063}, 2023.

\bibitem{guo2023sparsectrl}
Yuwei Guo, Ceyuan Yang, Anyi Rao, Maneesh Agrawala, Dahua Lin, and Bo~Dai.
\newblock Sparsectrl: Adding sparse controls to text-to-video diffusion models.
\newblock {\em arXiv preprint arXiv:2311.16933}, 2023.

\bibitem{Guajardo}
Jaime Guajardo, Ozgun Bursalioglu, and Dan~B Goldman.
\newblock Generative ai for 2d character animation.
\newblock In {\em ACM SIGGRAPH 2024 Posters}, SIGGRAPH '24, New York, NY, USA, 2024. Association for Computing Machinery.

\bibitem{Smith}
Harrison~Jesse Smith, Qingyuan Zheng, Yifei Li, Somya Jain, and Jessica~K. Hodgins.
\newblock A method for animating children’s drawings of the human figure.
\newblock {\em ACM Trans. Graph.}, 42(3), jun 2023.

\bibitem{kirillov2023segment}
Alexander Kirillov, Eric Mintun, Nikhila Ravi, Hanzi Mao, Chloe Rolland, Laura Gustafson, Tete Xiao, Spencer Whitehead, Alexander~C Berg, Wan-Yen Lo, et~al.
\newblock Segment anything.
\newblock In {\em Proceedings of the IEEE/CVF International Conference on Computer Vision}, pages 4015--4026, 2023.

\bibitem{sun2024}
Qirui Sun, Qiaoyang Luo, Yunyi Ni, and Haipeng Mi.
\newblock Text2ac: A framework for game-ready 2d agent character(ac) generation from natural language.
\newblock In {\em Extended Abstracts of the 2024 CHI Conference on Human Factors in Computing Systems}, CHI EA '24, New York, NY, USA, 2024. Association for Computing Machinery.

\bibitem{tevet2023human}
Guy Tevet, Sigal Raab, Brian Gordon, Yoni Shafir, Daniel Cohen-or, and Amit~Haim Bermano.
\newblock Human motion diffusion model.
\newblock In {\em The Eleventh International Conference on Learning Representations}, 2023.

\end{thebibliography}

\end{document}